\documentclass[aps,prb,reprint,superscriptaddress]{revtex4-1}

\usepackage{siunitx}
\usepackage[utf8]{inputenc}
\usepackage{graphicx}

\begin{document}

\title{Spin-dependent shot noise enhancement in a quantum dot} 

\author{Niels Ubbelohde}
\affiliation{Institut f\"ur Festk\"orperphysik, Leibniz Universit\"at Hannover, 30167 Hannover, Germany}
\author{Christian Fricke}
\affiliation{Institut f\"ur Festk\"orperphysik, Leibniz Universit\"at Hannover, 30167 Hannover, Germany}
\author{Frank Hohls}
\affiliation{Institut f\"ur Festk\"orperphysik, Leibniz Universit\"at Hannover, 30167 Hannover, Germany}
\affiliation{Physikalisch-Technische Bundesanstalt, 38116 Braunschweig, Germany}
\author{Rolf J. Haug}
\affiliation{Institut f\"ur Festk\"orperphysik, Leibniz Universit\"at Hannover, 30167 Hannover, Germany}

\date{\today}

\begin{abstract}
The spin-dependent dynamical blockade was investigated in a lateral quantum dot in a magnetic field. Spin-polarized edge channels in the two-dimensional leads and the spatial distribution of Landau orbitals in the dot modulate the tunnel coupling of the quantum dot level spectrum. In a measurement of the electron shot noise we observe a pattern of super-Poissonian noise which is correlated to the spin-dependent competition between different transport channels.
\end{abstract}


\maketitle

The prospect of using the electron spin as a basic element of information for quantum computation and new semiconductor devices has stimulated the study of spin-dependent phenomena. Many of these applications require  control of the spin-dependent dynamics in the transfer of charge through single electron devices. Fluctuations of the current contain information about these dynamics and the underlying mechanisms can be probed in a measurement of the electron shot noise.
In particular, the bunching of tunneling events, observable as an increased shot noise power, can characterize the Coulomb interaction in the transport through multilevel quantum dots \cite{Safonov2003, Onac2006,Sukhorukov2001, Thielmann2005, Zarchin2007, Zhang2007, Gustavsson2006, Barthold2006}. The sequence of tunneling events is correlated by the Coulomb blockade and depends on the effective tunneling rates and internal level structure, which allows for the detection of spin-dependent tunneling through quantum dots \cite{Cottet2004a}. Injection of electrons from spin-polarized leads may result in a spin-dependent blockade as shown by Ciorga et al. \cite{Ciorga2000,Ciorga2002,PioroLadriere2003}. The spin blockade effect \footnote{It should be noted that other spin selective mechanisms have been reported, which have a similar effect and are equally called spin blockade \cite{Weinmann1995,Ono2002}} has been observed in the addition spectrum of a quantum dot in a magnetic field \cite{Rogge2004,Rogge2006} as a modulation of the Coulomb blockade peak amplitude or in the occurrence of negative differential resistance \cite{Fricke2005}. While these measurements studied the average conductance, additional dynamical information can be obtained by shot noise detection techniques.

In this paper we present our measurements of the electron shot noise in the spin blockade regime. At finite bias we observe super-Poissonian noise at the Coulomb blockade peak. The shot noise enhancement follows a regular pattern correlated to the crossing Landau levels in a magnetic field. Interpretation in terms of the dynamical channel blockade mechanism suggests an underlying competition between transport channels with different spin.

\begin{figure}[!htbp]
\includegraphics{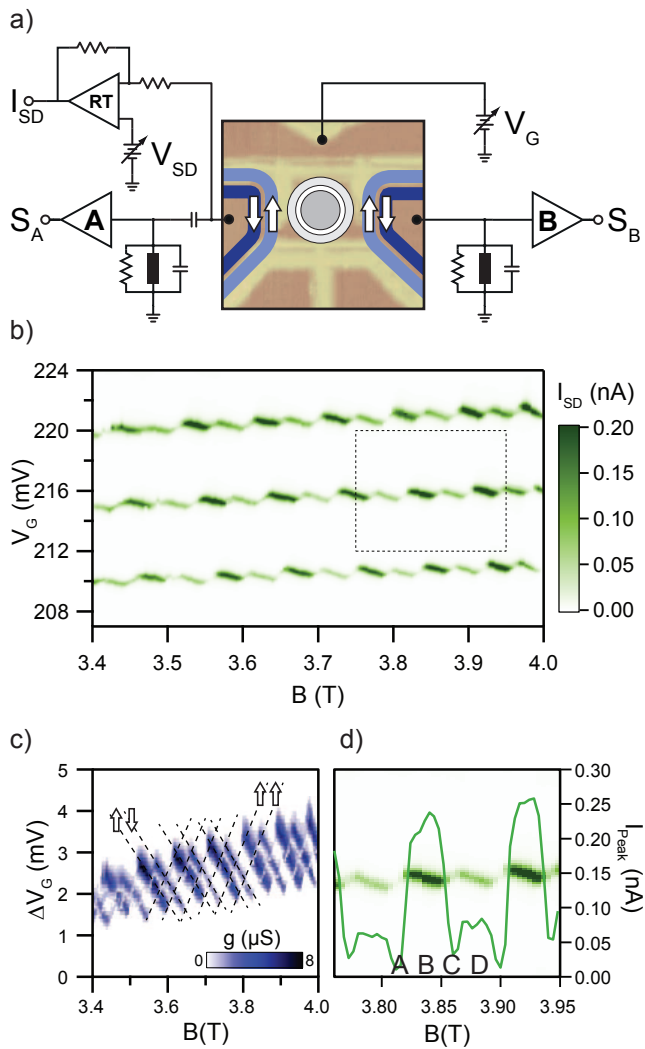}
\caption{\label{fig:fig1} (a) Simplified schematic diagram showing the low temperature amplifiers A and B connected to the RLC detector networks and the room temperature amplification (RT) of the DC transport signal. In magnetic field the spatial separation of the edge channels (blue) in the leads and the Landau orbitals (grey) in the quantum dot modulate the tunnel couplings. The AFM image of the sample structure shows the quantum dot formed by oxide lines(yellow). (b) Coulomb blockade peaks in the current for a quantum dot filling factor in the regime between 2 and 4 as a function of gate voltage and magnetic field. The dashed rectangle marks the regime investigated in more detail in (d) and Fig.~2. (c) The differential conductance $g = \frac{dI}{dV}$. The gaps between adjacent Coulomb blockade peaks are eliminated by removing a constant Coulomb repulsion energy \cite{Rogge2010} thereby shifting the upper and lower peaks visible in (b) towards the center peak. The lines indicate the crossing Landau levels and the regular pattern of spin pairing for the lowest Landau level.  (d) Spin blockade in the Coulomb blockade peak amplitude overlayed on a zoom of the image plot in (b).}%
\end{figure}

The quantum dot is defined by local anodic oxidation of a GaAs/AlGaAs heterostructure with a two dimensional electron system \SI{34}{nm} below the surface \cite{Fricke2005}. The electron density of the heterostructure is \SI{4.6e11}{cm^{-2}} and the mobility is \SI{6.4e5}{cm^2/V.s}. Measurements were performed in a $^3$He/$^4$He dilution refrigerator at a temperature of \SI{100}{mK}. Current fluctuations are converted to voltage fluctuations by RLC-circuits in a cross-correlation configuration (Fig.~\ref{fig:fig1}.a) centered at \SI{1}{MHz}. The voltage fluctuations are amplified by cryogenic HEMTs \cite{DiCarlo2006a, Ubbelohde2012} and digitized at room temperature. The real part of the cross-correlation noise power is evaluated within a frequency window from \SI{500}{kHz} to \SI{3}{MHz}. The DC-part of the source drain current $I_{SD}$ is measured with a transimpedance amplifier, which also biases the sample. The Coulomb diamond structure yields a charging energy of \SI{\sim 0.7}{meV} and an excitation energy of \SI{\sim 0.1}{meV}. From the total capacitive coupling and spatial extension of the dot a total charge of around 100 electrons is estimated.

Figure~\ref{fig:fig1}.b shows the current through the quantum dot as a function of gate voltage and magnetic field at a constant small bias (\SI{\sim0.02}{mV}). The chosen energy and magnetic field range spans over several Coulomb blockade peaks with the filling factor of the quantum dot being in the regime between 2 and 4. The peak energy position follows the well known zigzag pattern corresponding to the crossing of the first and second Landau level \cite{Fricke2005,Fuhrer2001,McEuen1991,McEuen1992}. The energies of the states in the lowest Landau level drop as the magnetic field is increased while the energies of the states in the second Landau level rise in this magnetic field range.

Figure~\ref{fig:fig1}.c shows the differential conductance as a function of gate voltage and magnetic field. The constant gaps between adjacent Coulomb peaks due to the Coulomb repulsion energy are removed following the procedure presented in detail in Ref. \cite{Rogge2010}. The states of the excitation spectrum can thereby be followed over several electron numbers. Lines with a positive slope connecting the onset of transport indicate the second Landau level. Lines with a negative slope following the peaks in the differential conductance highlight the energy and magnetic field dependence of the first Landau level. The lines belonging to transport through the first Landau level appear in pairs corresponding to the occupation of the same Landau orbital with opposite spin. The energy-level spacing for the second Landau level is however more regular and lacks the spin-pairing indicating different spin configurations for the first and second Landau level and an interaction induced spin polarization \cite{Rogge2010} of the higher Landau level.

Not only the peak position but also the peak amplitude (Fig.~\ref{fig:fig1}.d) is strongly influenced by the crossing of the two lowest Landau levels. The spatial separation between states in the inner and outer Landau orbitals results in unequal tunnel coupling to the leads. This is observable in the current as a reduced peak amplitude for states belonging to the second (inner) Landau level \cite{McEuen1991}. The corresponding upward slopes in Fig.~\ref{fig:fig1}.d are noticeably less pronounced. Additionally, the peak amplitude also alternates for electrons with opposite spin transferred through the lowest Landau level due to the spin-polarized edge states in the two-dimensional leads \cite{Ciorga2002,Ciorga2000}. The combination of this spin blockade effect and the spatial overlap with two Landau orbitals therefore gives rise to four regions (A to D in Fig.~\ref{fig:fig1}.d) with different peak amplitudes \cite{Ciorga2000}.

\begin{figure}
\includegraphics{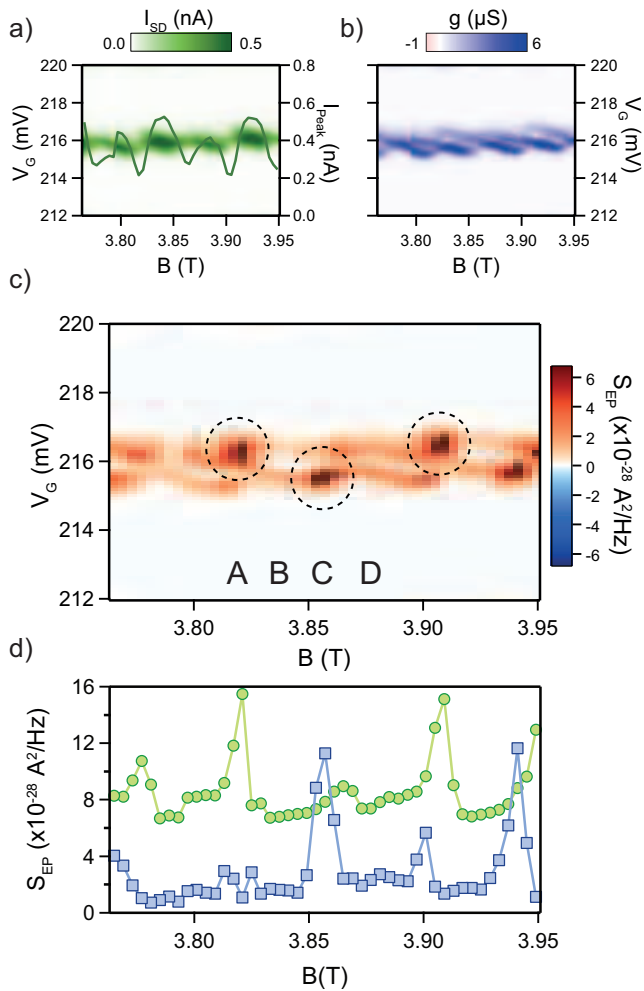}
\caption{\label{fig:fig2}Current (including peak amplitude) (a) and differential conductance (b) in the non-linear regime. (c) Excess noise, i.e. the difference of the measured noise power and the noise power of a Poissonian noise source with equal average current. Red indicates super-Poissonian noise. The dashed circles mark the peaks occurring alternately at the emitter and collector resonance. (d) Peak amplitude of the two super-Poissonian stripes. The amplitude for the stripe at larger gate voltages (green, circles) is shifted by \SI{6e-28}{A^2/Hz} with respect to the stripe at lower gate voltages (blue, squares).}%
\end{figure}

In this system with spin-dependent tunnel couplings the fluctuations of the current are analyzed at a bias of \SI{0.2}{mV}, where the Coulomb blockade peak is broadened into a current stripe (Fig.~\ref{fig:fig2}.a). Therefore an additional transport channel of the excitation spectrum is able to enter the transport window defined by the bias between emitter and collector lead. Spin blockade still suppresses (although less pronounced) transport and the modulation of the peak amplitude remains visible in the current. The differential conductance in Fig.~\ref{fig:fig2}.b shows two stripes corresponding to the rising and falling edges of the Coulomb blockade peak and indicates the two resonances with the emitter and collector lead.

Figure~\ref{fig:fig2}.c shows the analyzed current fluctuations in the form of the excess noise, which is the difference between the measured shot noise power and the expectation value of a Poissonian noise source of equal average current. In comparison to the often used Fano factor the excess noise is less susceptible to divergence, if transport is blocked and the current is small \cite{Zhang2007}, but still offers a qualitative picture of the presence of additional correlations.

The current fluctuations show two stripes of super-Poissonian noise, resembling the pattern observed in the differential conductance. Depending on the gate voltage multiple transport channels of the excitation spectrum are available within the transport window. The energy-dependent occupation probability in the leads modulates transport through the level spectrum. The resulting differences in the occupation life time lead to dynamically occurring discontinuities in the sequence of transferred charges as the charging energy prohibits the tunneling of more than one electron at a time. This bunching of tunneling events results in an increased noise power compared to the Poissonian noise of single barrier tunneling. Experiments with multilevel quantum dots demonstrated how this dynamical channel blockade mechanism gives rise to super-Poissonian shot noise at both the emitter and collector resonance of a Coulomb blockade peak \cite{Zhang2007, Zarchin2007}. Zarchin et al. \cite{Zarchin2007} found the shot noise to be strongly enhanced in a magnetic field, while the origin of the multilevel system could not be clearly identified. For our system and the magnetic field range presented in Fig.~\ref{fig:fig2} the multilevel system depends on the occupation of the two Landau levels and the different spin levels. The stripes of super-Poissonian noise in Fig.~\ref{fig:fig2}.c follow the magnetic field dependence of the crossing Landau levels allowing for the identification of four regions very similar to A to D in Fig.~\ref{fig:fig1}.d.

The most pronounced super-Poissonian peaks in the excess noise are detected in the regions corresponding to A and C, where at small bias, charge is transferred through the inner Landau orbital and transport is strongly suppressed. In comparison, shot noise therefore offers a complementary image to the current and can indicate the competition between blocked and unblocked transport channels, even if this effect cannot be observed as negative differential conductance (Fig.~\ref{fig:fig2}.b) \cite{Belzig2005d}, which has previously been used to detect spin blockade in the nonlinear regime \cite{Ciorga2002}.

The peaks in the excess noise appear in a regular pattern. For consecutive regions with transport through the inner Landau orbital the peak position alternates between the emitter and collector resonance (Fig. \ref{fig:fig2}.c). In A the peak appears at the collector resonance and in C at the emitter resonance. This is in contrast to experiments performed at zero magnetic field \cite{Zhang2007}, where the super-Poissonian noise appears nearly symmetric to the Coulomb blockade peak.

\begin{figure}
\includegraphics{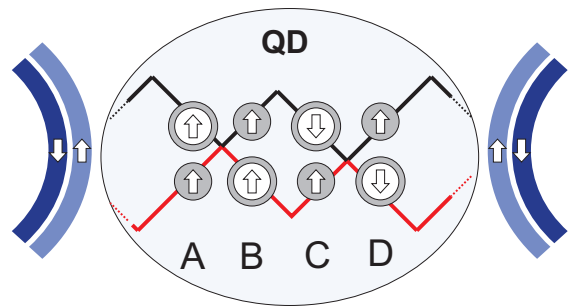}
\caption{\label{fig:fig3}Simplified energy diagram of the basic multilevel system in a magnetic field consisting of a ground state (red) and excited state (black). Arrows symbolize the spin of the tunneling electron, while the circles depict the transfer into the first, inner Landau orbital (small filled circle) and the second, outer Landau orbital (open circle with a filled ring). In total there are therefore four different configurations (A to D) for the multilevel system. The varying tunnel coupling to the spin-polarized leads (blue) results in a dynamical blockade between excited and ground state.}%
\end{figure}

As indicated by the regular energy level spacing the occupation of the second Landau level is assumed to be spin-polarized while the transport through the first Landau-level changes spin, confirmed by the peak amplitude in DC-transport. The spin configuration and energy spacing of the quantum dot level system depends on magnetic field and influences the mutual dynamical blockade between ground and excited state in the transport window. The resulting four configurations of the basic multilevel system correspond to the four regions A to D identified in the experiment, as illustrated in a simplified energy diagram in Fig.~\ref{fig:fig3}. In the case of the super-Poissonian peaks measured in region A and C, the second Landau level constitutes the spin-polarized ground state while the first Landau level forms the excited state. The regions A and C have opposite spin states. In B and D the level configuration is reversed, with the first Landau level as the ground state and the second Landau level as the excited state. In all cases both states differ in their respective coupling to the spin-polarized leads. There is a peak in the excess noise, if the weakly coupled second Landau level forms the ground state. The position of these peaks changes with the spin of the excited state. The relative contribution of these channels towards transport through the emitter and collector barrier further depends on the occupation probability in the leads, which modifies the difference in the effective tunneling rates at the corresponding lead-resonance. Following these basic considerations the periodic pattern of the observed shot noise enhancement, occurring most visibly at only one lead-resonance, is evidence of the alternating spin configuration.

The occurrence of super-Poissonian noise has been modeled theoretically for various quantum dot circuits \cite{Cottet2004a, Cottet2004, Weymann2007}, taking into account, for example, the spin polarization of ferromagnetic leads or the influence of a magnetic field. In all these cases the Coulomb blockade mechanism is fundamental for the shot noise enhancement. Cottet et al. \cite{Cottet2004} also briefly discuss a dynamical spin blockade for the more closely related case of the magnetic field applied not only to the dot locally but to the whole circuit including the leads and approximate it with a Zeeman-split dot level with spin-polarized leads. The spin blockade mechanism demonstrated in this simplified system supports spin-dependent tunneling rates as the origin of the observed shot noise enhancement. However, in the experiment the system is further complicated by the different coupling of the Landau-orbitals in the quantum dot and the magnetic field dependence of the energy level spacing.

In summary, we have investigated the current fluctuations in a quantum dot with spin-polarized leads. We observe a regular pattern of super-Poissonian shot noise corresponding to position and amplitude modulation of the well known zigzag pattern in DC transport. Based on previous experimental results \cite{Zhang2007, Zarchin2007, Gustavsson2006} the complex internal level structure in this regime implies a dynamical channel blockade as the mechanism behind the shot noise enhancement. The observed alternating pattern shows a dependence on spin in the studied magnetic field range. The experiment thus demonstrates the detection of spin-dependent dynamics in quantum dot systems in a measurement of the electron shot noise.

\begin{acknowledgments}
The authors gratefully acknowledge financial support by the Deutsche Forschungsgemeinschaft and the Centre for Quantum Engineering and Space Time
Research (QUEST). We also thank W. Wegscheider for providing the wafer material, and B. Harke for fabricating the device.
\end{acknowledgments}

%

\end{document}